\documentclass[prl,twocolumn,showpacs]{revtex4}
\usepackage{graphicx}
\usepackage{amsmath}
\usepackage{amssymb}
\usepackage{color}
\usepackage[normalem]{ulem}
\usepackage{epsfig}
\usepackage{epstopdf}
\usepackage{latexsym}
\usepackage{url}
\usepackage{graphicx}
\usepackage{tabu}
\usepackage{bmpsize}
\usepackage[utf8]{inputenc}
\usepackage{setspace}

\begin{document}
\title{Predicting catastrophic shifts}
\author{Haim Weissmann and Nadav M. Shnerb}
\affiliation{Department of Physics, Bar-Ilan University, Ramat-Gan
IL52900, Israel} 
\begin{abstract}
Catastrophic transitions, where a system shifts abruptly between
alternate steady states, are a generic feature of many nonlinear
systems. Recently these regime shift were suggested as the mechanism
underlies many ecological catastrophes, such as desertification and
coral reef collapses, which are considered as a prominent threat to
sustainability and to the well-being of millions. Still, the methods
proposed so far for the prediction of an imminent transition are
quite ineffective, and some empirical and theoretical studies
suggest that actual transitions may occur smoothly, without an
abrupt shift. Here we present a new diagnostic tool, based on
monitoring the dynamics of clusters through time. Our technique
discriminates between systems with local positive feedback, where
the transition is abrupt, and systems with negative density
dependence, where the transition is smooth. Analyzing the spatial
dynamics of these two generic scenarios, we show that changes in the
critical cluster size provide a reliable early warning indicator for
both transitions. Our method may allow for the prediction, and thus
hopefully  the prevention of such transitions, avoiding their
destructive outcomes.
\end{abstract}
\maketitle

%
%
%
%
%
%
%

\section{Introduction}

The stability of ecosystems, and in particular the response of
populations and communities to external perturbations, is one of the
main topics in contemporary science \cite{muller2010resilience}. As
the impact of anthropogenic changes (carbon emission, habitat
fragmentation, introduction of non-indigenous species and pathogens)
reaches the global scale, worries about their potential outcomes are
growing \cite{dawson2011beyond}. Recently, there is an increasing
concern about the scenario known as catastrophic regime shift, where
a relatively small change in the environmental conditions leads to a
sudden jump of the system from one state to another
\cite{scheffer2001catastrophic,scheffer2012anticipating}. This
change is often irreversible and accompanied by hysteresis: once the
system relaxes to its new state, it will not recover even when the
environmental conditions are restored.

\begin{figure*}
\begin{center}
\includegraphics[width=18cm]{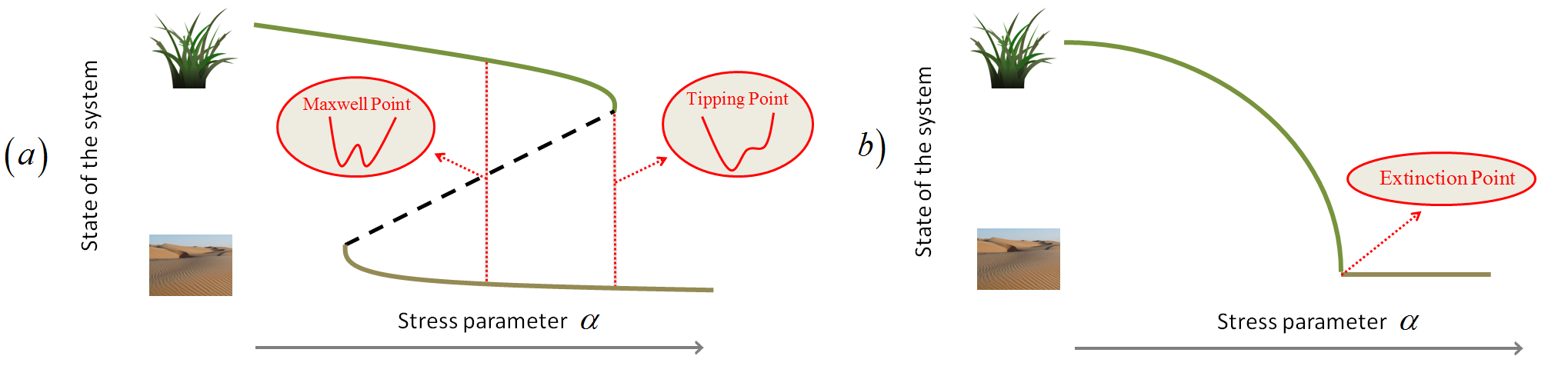}
\vspace{-0.8cm}
\end{center}
\caption{\textbf{Catastrophic shift vs. continuous transition.} The
generic features of a nonlinear system that supports catastrophic
shift are illustrated in panel a (left). The two stable state (full
lines, here one represents vegetation, the other bare soil) coexist
for some region of the stress parameter $\alpha$. The transition may
take place at the tipping point (right dotted line), where the basin
of attraction of the vegetation state (corresponding to the right
well in the circled cartoon) vanishes, and its attractiveness (the
curvature of the well) approaches zero. In spatial systems, on the
other hand, large bare-soil clusters will invade vegetation to the
right of the Maxwell point (left dotted line), where the stability
of both alternate states becomes equal. Under disturbances, the
transition takes place at the MP \cite{bel2012gradual}. A continuous
transition scenario is illustrated in panel b (right), where
vegetation went extinct as the stress is growing. The theory of
extinction transitions of this type also suggest diverging
spatio-temporal fluctuations at the transition point
\cite{hinrichsen2000non}.}
 \label{bifurcation}
\end{figure*}

One of the main topics considered in the context of catastrophic
shifts is the possibility of sudden extinction of populations as the
environment varies
\cite{drake2010early,takimoto2009early,peters2012directional}. For
example, changes in solar radiation owing to variations of the
Earth's orbit may have triggered the sudden mid-Holocene (5000~yrs
ago) desertification of the Sahara \cite{scheffer2001catastrophic}.
The standard model used to describe this phenomenon involves
nonlinear dynamics that supports two alternate steady states  with a
(backward) fold bifurcation
\cite{scheffer2001catastrophic,rietkerk1997alternate}. This
mechanism is illustrated in Figure \ref{bifurcation}a: if the
parameter $\alpha$ stands for environmental \emph{stress} (e.g.
grazing, or decreased precipitation) the system supports, for
certain values of $\alpha$, two stable states, one corresponds to
vegetation and the another to bare soil. This bistability is related
to the nonlinearity of the system and reflects a positive feedback
mechanism
\cite{hillerislambers2001vegetation,holmgren1997interplay}, such
that vegetation grow above some critical density, while below this
density the vegetation declines.

Following this insight, the search for early warning indicators that
will allow one to predict an imminent transition has become a major
research topic in the last decade.  Most of these efforts have been
focused on the phenomenon known as critical slowing down, meaning
that one of the stable states (say, the vegetation state in Fig.
\ref{bifurcation}a) looses its stability at the tipping point
\cite{eslami2014approaching,scheffer2009early}. Accordingly, the
rates at which the system recovers from spatial or temporal
perturbations become slower and slower as it approaches the
catastrophe. This feature has, indeed, been demonstrated in recent
experiments
(e.g.\cite{drake2010early,dai2013slower,veraart2012recovery,carpenter2011early}).

However, a few new  studies cast a severe doubt regarding the
relevance of these indicators  to empirical ecological dynamics.
First, critical slowing down and its consequences, like fat tailed
or skewed patch statistics, does not necessarily indicate a tipping
point or discontinuous transition. All these features are also
characteristic of continuous transitions, where the system changes
its state smoothly and reversibly without hysteresis
\cite{kefi2013early,eslami2014approaching}. A schematic illustration
for such a scenario is given in Fig. \ref{bifurcation}b, as an
increase in stress leads to gradual extinction without bistability.
Continuous transitions of this type characterize various ecological
models that describe generic processes, including logistic growth
without an Alley effect and the susceptible-infected-susceptible
(SIS) model for epidemics. In both cases, and under many other
dynamics, the transition to extinction as the birth/infection rate
decreases is continuous with no sudden jumps, yet the response of
the system to external perturbations becomes infinitely slow close
to the transition point (see, e.g.,
\cite{kessler2007solution,kessler2008epidemic}). A few recent
studies, showing a recovery from desertification when the external
pressure (grazing, in most cases) has been removed
\cite{fuhlendorf2001herbaceous,rasmussen2001desertification,valone2002timescale,zhang2005grassland,Allington2010973},
suggest also that the transition is, at least in some cases,
continuous and reversible.

Another line of criticism has to do with spatial structure. When a
system admits two stable states, local disturbances and fluctuations
often generate patches of an alternate state, like regions of bare
soil surrounded by vegetation and vice versa. As  discussed  in more
detail below, the Maxwell point (MP, shown in Fig.
\ref{bifurcation}a) marks the boundary between two  regimes: to the
right of the MP, large patches of bare soil invade vegetation, while
to the left of the MP vegetation invades bare soil
\cite{durrett1994importance} (MP is also known as the melting point
in the theory of first order transitions, and it is the stall point
for a front connecting two metastable states). Accordingly, for the
generic case of spatial system with stochastic dynamics one should
expect the transition to take place close to the Maxwell point, not
at the tipping point \cite{bel2012gradual}. However, at the Maxwell
point both states are stable, as seen in Fig. \ref{bifurcation}a,
and there is no critical slowing down. Accordingly,  all the early
warning criteria that are based on the slow recovery of the system
at the vicinity of the tipping point will fail to predict the
crossing of the Maxwell point, which is the relevant factor that
drives the catastrophe in this typical scenario.

Here we would like to suggest a new method aimed at identifying the
state of the system. Our method both distinguishes between
continuous transitions and catastrophic shifts and provides a
quantitative measure of the distance from the transition. This
method is based on the monitoring of the cluster dynamics, and in
particular the probability of a cluster to grow or shrink as a
function of its size. It turns out that this technique reveals the
nature of the transition: a catastrophic shift is characterized by a
positive correlation between cluster size and its chance of growing,
while for a continuous transition the opposite is true, as small
clusters tend to grow where as large clusters shrink. The distance
from the transition, in both cases, is related to the critical
cluster size, and we will show that, as the system approaches the
transition point, this size diverges for discontinuous transitions
and goes to zero for continuous transitions.

This work, as we shall explain below, is based on simple insights
gained from nucleation theory (for discontinuous transitions) and
the theory of extinction dynamics. To demonstrate its power, we
present a numerical study of  two generic models  - the
Ginzburg-Landau model of irreversible transitions and the contact
process model for gradual extinction. Both models are analyzed in
the context of desertification, i.e., a transition from vegetation
to bare-soil state. However the main lessons acquired are relevant,
\emph{mutatis mutandis}, to the analysis of catastrophic and
non-catastrophic transitions in general.

\section{Results}

Our cluster dynamics method is, of course, strongly related to the
fact that our system is spatial \cite{durrett1994importance}. The
spatial dynamics of populations is usually modeled by some kind of
''diffusion" term, representing the random movement (or dispersal)
of individuals among neighboring patches. More generally, the
feature that lies behind the results presented here is that the
spatial dynamics prefers ''smeared" spatial patterns, i.e., the
system is trying to avoid strong spatial gradients of population
density.

As mentioned above, the main characteristic of bistable systems that
allow for catastrophic transitions is \emph{positive feedback}: in a
local patch, small populations go extinct and large populations are
self-sustained. However, when  the spatial dynamics is taken into
account, small patches, for which the area of the surface region is
large with respect to their ''volume", are under stronger stress
from their neighborhood, while the effect of surface stress is
vanishingly small for large patches. This phenomenon is analogous to
the opposing effects of surface tension and bulk free energy that
governs the physics of nucleation in first order transitions
\cite{kelton1991crystal}. As a result, one expects that, for the
same value of the external parameter (say, $\alpha$), large clusters
are more stable, and their tendency to grow (or at least not to
shrink) is enhanched with respect to small clusters.

When the system has no, or very weak, positive feedback, there is no
bistability, and population density goes continuously to zero at
some critical value of the external parameter. However, as discussed
by many authors (e.g., \cite{bonachela2012patchiness} and references
therein), the spatial structure of the system is still very
relevant. The fate of a population depends on the ratio between
birth and death rates. In spatial systems the local \emph{negative}
feedback (meaning that individuals cannot reproduce in, or into,
occupied sites), when superimposed on the emergent clustering (since
death occurs everywhere but reproduction is local) leads to a
decrease of the effective birth rate. As a result, the transition in
spatial system takes place when the per capita birth rate is larger
than the death rate. Accordingly, for these systems the effect is
just the opposite: the smaller the cluster, and the larger its
interface with empty (or low density) sites, the larger its chance
to grow.

\begin{figure*}
\begin{center}
\includegraphics[width=18.5cm]{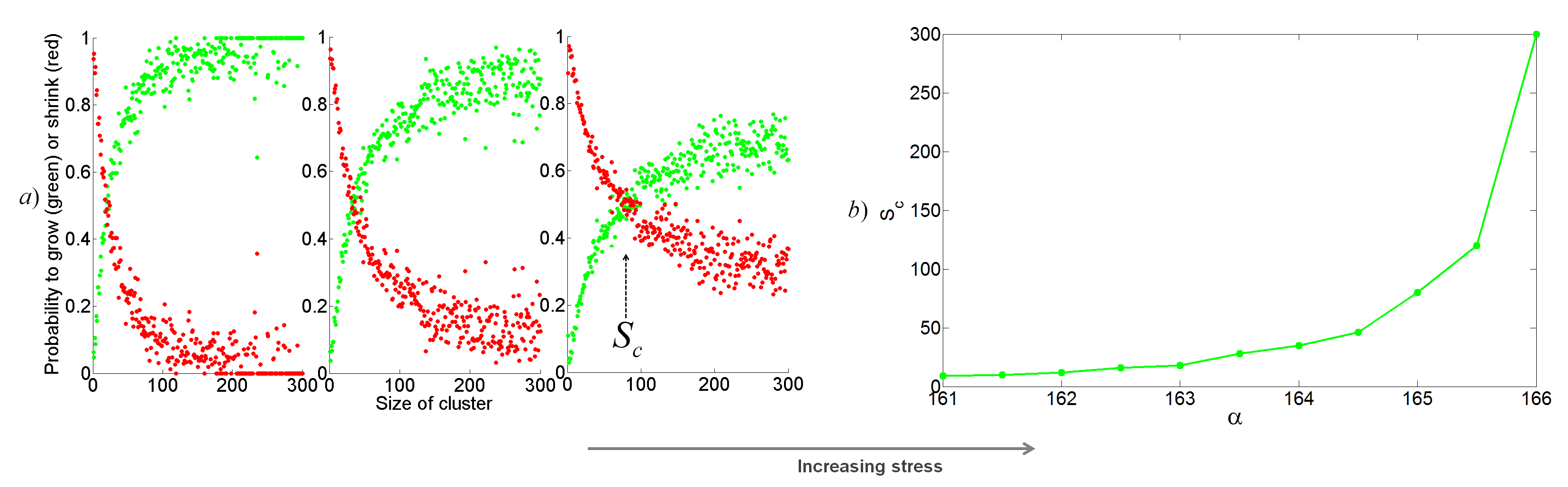}
\vspace{-0.8cm}
\end{center}
\caption{\textbf{Cluster dynamics in bistable system.} Monitoring
the evolution of clusters in a bistable Ginzburg-Landau system with
environmental noise (see methods), the chance of a cluster to grow
(green) or to shrink (red) is plotted against its size for various
values of $\alpha$: $163$ (panel a, left), $164$ (middle) and $165$
(right). Clearly, the chance of a cluster to grow  is positively
correlated to its area. As the environmental conditions deteriorate
the minimal size of a growing cluster is increasing, so the value of
$S_c$ grows with $\alpha$.  In panel (b) the $S_c$ is plotted
against $\alpha$, with apparent divergence as $\alpha$ approaches
$\alpha_{MP}$.  The parameters for the figures $\zeta=2$,
$\eta=1000$, $D=6$ and $S=100\times100$.} \label{catastrophe}
\end{figure*}

\begin{figure*}
\begin{center}
\includegraphics[width=18cm]{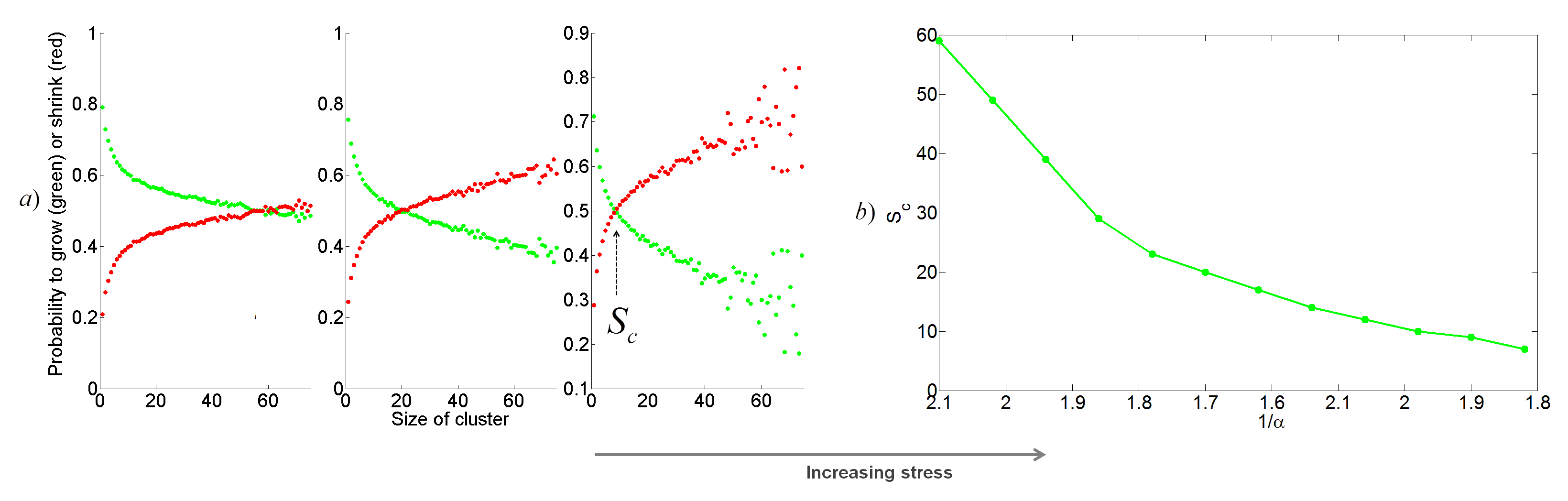}
\vspace{-0.8cm}
\end{center}
\caption{\textbf{Cluster dynamics for a system approaching a
continuous transition} The time evolution of vegetation clusters
described by a contact process (birth-death process with one
individual per site, see methods) was monitored. In contrast with
the behavior illustrated in Fig \ref{catastrophe}, here the chance
of a cluster to grow (green) increases when its size decreases
(panel a). The three subplots correspond to $\alpha$: $0.47$ (left),
 $0.52$ (middle) and $0.58$ (right). Moreover, as the
stress increases, $S_c$ decreases, as only individuals surrounded by
bare soil admit positive growth rate (panel b). Results were taken
from simulation on a $100\times100$ lattice with $\zeta=2$,
$\eta=1000$ and $D=6$.} \label{contact}
\end{figure*}

To demonstrate the applicability of these qualitative insights, we
have analyzed two generic models. For the case of catastrophic
shift, we used, as in \cite{bel2012gradual,weissmann2014stochastic}
the Ginzburg-Landau model, which is the simplest nonlinear dynamics
that provides both positive feedback and finite carrying capacity.
To model a continuous extinction transition we have implemented the
contact process, a canonical model of a birth-death process on
spatial domains. As suggested by Grassberger and Jansen
\cite{janssen1981nonequilibrium,grassberger1982phase}, continuous
extinction transitions belong generically to the directed
percolation equivalence class, for which the contact process is a
standard example. In the context of population dynamics, the
applicability of this conjecture was demonstrated recently by
\cite{bonachela2012patchiness}),

In Figure \ref{catastrophe} the chance of a cluster to grow/shrink
is plotted against its size for the Ginzburg-Landau model (see
methods). Clearly, the larger  the size of a cluster, the larger is
its chance to grow. Just the opposite feature is demonstrated in
Fig. \ref{contact} for the continuous transition: here the chance of
a cluster to grow is negatively correlated with its size. This
qualitative feature is quite prominent and may allow one to identify
the nature of the system (bistable or not) and to guess the
characteristics of an imminent transition (continuous or
catastrophic) even with poor-quality data.

A second feature demonstrated in Figs. \ref{catastrophe} and
\ref{contact} is the appearance of a critical cluster size $S_c
(\alpha)$. Clusters of size $S_c$  neither shrink nor grow on
average. In catastrophic (positive feedback) systems smaller
clusters shrink and larger cluster grow, while the opposite is true
when the transition is continuous. As the value of $\alpha$
approaches the Maxwell point ($\alpha_{MP}$) for a bistable system,
$S_c \to \infty$, meaning that no vegetation cluster grows on
average above $\alpha_{MP}$. On the other hand, in a continuous
transition $S_c$ takes its minimal value at the extinction point,
indicating that even small clusters cannot grow anymore.

Accordingly, our suggested diagnostic procedure has two stages. The
first is based on (at least) two snapshots of the spatial system,
allowing for a comparison of the chance of a patch to shrink or to
grow, thus indicating the type (bistable/catastrophic or
monostable/continuous) of the system. Comparing (at least) three
snapshots and tracing the value of $S_c$ along time one obtains an
early warning indication of an imminent transition if $S_c$ grows
(in a catastrophic system) or shrinks (in a continuous system). This
procedure is summarized in  table \ref{table1}.

\begin{table}
\begin{center}
\includegraphics[width=7cm]{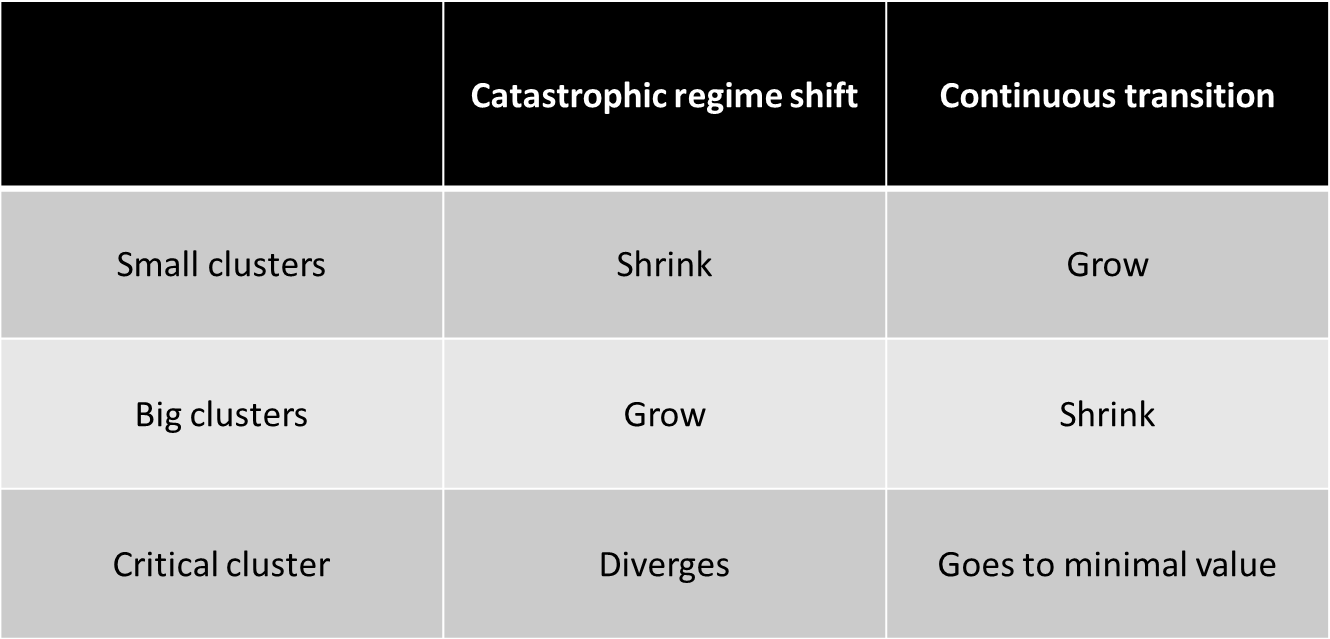}
\caption{Summary of the suggested indicators.} \label{table1}
\end{center}
\end{table}

If a bistable system is in its extinction phase  (i.e., $\alpha >
\alpha_{MP}$), where a large enough patch of bare soil will invade
vegetation, but the disturbance that creates this void has not yet
happened) the small patch dynamics still provides an indication as
to the state of the system, as the lines representing the chance to
grow/shrink (see Fig. \ref{catastrophe}) will level off and
saturate, indicating that $S_c \to \infty$ and that the system is
living on  borrowed time.


\section{Discussion}

The main insight we have implemented in this paper is the
distinction between systems with positive feedback, i.e, positive
correlation between the fitness of individuals and the density, and
systems with only negative feedback, where an increase of the local
density is followed by a decrease of individual's fitness. In the
first case one expects an alternate steady state, hysteresis and
catastrophic shift, while in the other case the extinction
transition is continuous. As explained above, positive feedback
leads to an access growth of large clusters, while in systems with
negative feedback small clusters are favorable.

Accordingly, we have suggested two diagnostic tools, both are based
on comparison between  consecutive snapshots taken from the same
spatial domain. First, by comparing (at least) two snapshots one may
obtain a quantitative assessment of the importance of positive
feedback by measuring the correlation between the size of a cluster
and its chance to grow or shrink. Using (at least) three snapshots
one may get an early indication of an immanent transition, as $S_c$
diverges (in the catastrophic shift scenario) or shrinks to low
values (in the continuous transition case).

Of course, systems with pure positive feedback or pure negative
feedback are just the two extremes of a continuum. In many systems
positive and negative feedback interfere, and their relative weights
determine the characteristics of the transition. In particular,
Ginzburg-Landau systems with demographic stochasticity yield a
continuous (negative feedback controlled) transition in  one spatial
dimension \cite{weissmann2014stochastic}, and switch to a positive
feedback bistable transition for weak demographic noise in two
dimensions \cite{martin2015eluding}. Still, when considered from the
cluster dynamics perspective, the only factor that determines the
nature of the transition is the feasibility of an invading cluster,
and our analysis addresses precisely this point.

Another interesting scenario is the case of a \emph{neutral}
dynamics \cite{Hubbell2001unifiedNeutral}, when there are no
deterministic forces and no attractive fixed point, and the
evolution of the system is governed solely by demographic and
environmental stochasticity. Under neutral dynamics the chance of a
cluster to grow or shrink is independent of its size. Accordingly,
when the figures that correspond to \ref{catastrophe} and
\ref{contact} show a straight line, one may deduce that the dynamics
(at least up to the relevant length and time scales) is neutral.
This feature was, indeed, demonstrated in \cite{seri2012neutral}
(Fig. 3) for clusters of trees in the tropical forest.

In summary, we have suggested a general diagnostic tool that may
serve any specific study of a potential transition on spatial
domain. Tracking cluster dynamics along a certain period reveals the
dominant mechanism (positive/negative feedback) that governs the
dynamics, the expected character of a transition (smooth/abrupt) and
its proximity. We hope that this technique will enhance the
predictive ability of relevant studies, assisting the effort to
avoid undesirable catastrophic transitions, together with  their
disastrous consequences.

\section{Methods}
Along this paper we consider and simulate two generic models, one
that supports a catastrophic shift and irreversible transition, and
another that give rise to a continuous transition without
hysteresis. Here we describe the models and our simulation
technique.

\textbf{Catastrophic transition:} We have implemented the
Ginzburg-Landau model, which is the minimal model that describes a
discontinuous (first order) transition. In the context of
desertification we are looking at the biomass density, $b$, which
satisfies:
\begin{eqnarray} \label{GL}
\frac{\partial b}{\partial t}=D\nabla^2 b-\alpha b+\beta
b^{2}-\gamma b^{3}.
\end{eqnarray}
Here the $\beta$ term models the effect of positive feedback (an
increase of the growth rate with density) and the $\gamma$ term
enforces a finite carrying capacity. The diffusion term reflects the
spatial spread of the biomass, e.g., plant dispersal. For further
details, see \cite{weissmann2014stochastic}.

In this model, $b=0$ is the bare soil state and the alternative
uniform solution $b = -\alpha/(2 \gamma) + \sqrt{\beta^2 - 4 \alpha
\gamma}/(2 \gamma)$ is the state with vegetation. An increase in the
control parameter $\alpha$ corresponds to increased stress (less
precipitation, more grazing etc.). Beyond the tipping point at
$\alpha_{TP}=\beta^2/(4 \gamma)$ the vegetation state no longer
exists; as $\alpha$ crosses $\alpha_{TP}$, a catastrophic shift
occurs and the system collapses to the bare soil state. To restore
the vegetation state the strength of the environmental pressure has
to be reduced until it passes through the other tipping point at
$\alpha =0$.

Simulations of this model were preformed on a $2d$, $S=L\times{L}$
lattice with periodic boundary conditions. The deterministic
dynamics was simulated via Euler integration of (Eq. \ref{GL}) with
$\Delta t = 0.001$, implementing asynchronous update to avoid
artifacts like fictitious bias of the dispersal. To add disturbances
to the model, after every time interval $\tau$  the biomass at each
site was multiplied by $1+\eta$, where $\eta$ is a random number
taken from a uniform distribution between $-\Delta$ and $\Delta$.
The parameters used for the results presented in Fig.
\ref{catastrophe} are $\beta=40, \  \gamma=1.6$ (hence
$\alpha_{MP}=222.22$ and $\alpha_{TP}=250$), $L=100, \ \tau =2, \
\Delta = 0.4$

\textbf{Contact process extinction transition (Lattice SIS):} In a
contact process every site is either empty or occupied by one
individual (active). An active site dies at a rate one, and is
trying to reproduce at rate $1/\alpha$ (again, $\alpha$ is a stress
parameter, an increase in $\alpha$ leads to a decrease of the birth
rate). When an individual tries to reproduce, it picks at random one
of its neighboring sites, and if the chosen site is empty, it
becomes active, otherwise, nothing happens. Accordingly, the
productivity of a site is inversely proportional to the local
density.

The process was simulated using the Gillespie algorithm on a $2D$,
$100 \times 100$ lattice. It is known
\cite{dickman1998moment,hinrichsen2000non} that in this case the
transition is continuous and extinction takes place at $\alpha_c
\approx 0.61$.

\textbf{Cluster tracking} Trying to emulate the results of
consecutive censuses of an empirical system, we have "sampled" our
system  every $p$ generations: a snapshot of the spatial pattern was
taken and the dynamics of clusters is obtained by comparison with
the previous snapshot. For the contact process, the definition of a
spatial cluster is trivial: it is a collection of active sites in
which every pair is connected by a path of nearest neighbor active
sites. For the catastrophic shift model every site was classified as
"active" if $b_{i,j}$ is in the basin of attraction of the
vegetation state and as "inactive" if this point is attracted to the
bare soil state. These clusters were identified and labeled using
MATLAB's \textit{bwlabel} subroutine.

To track the evolution of clusters we have implemented a simple
motion tracking algorithm (see, e.g.,
\cite{seri2012neutral,falkowski2006mining,hartmann2014clustering},
and then the cluster at one snapshot is compared with the previous
one to identify growth or decay.


\bibliography{references_indicator}

\end{document}